\documentclass[11pt,a4paper]{article}
\usepackage{amsmath,amssymb,amsfonts,a4wide,graphicx,bm,times,mathtools} 
\usepackage{mcite}

\newcommand{\daniele}[1]{$\framebox{\tiny DG}$\ \textbf{\texttt{{\color{Orange}\footnotesize#1}}}}

\usepackage{hyperref}
\hypersetup{
    colorlinks,
    citecolor=green,
    filecolor=green,
    linkcolor=blue,
    urlcolor=green
}

\makeatletter \let\old@startsection=\@startsection
\renewcommand{\@startsection}[6]
{\old@startsection{#1}{#2}{#3}{#4}{#5}{#6\mathversion{bold}}}
\makeatother

\makeatletter

\@addtoreset{equation}{section}
\makeatother

\newcommand{\be}{\begin{equation}}
\newcommand{\ee}{\end{equation}}
 \def\i{\mathrm{i}}

 \def\Re{{\rm Re ~}}
 \def\Im{{\rm Im ~}}

\def\CF{{\mathcal{F}}}

\def\CN{{\mathcal{N}}}

\def\CP{{\mathcal{P}}}

\usepackage[english] {babel}
\usepackage [latin1]{inputenc} 
\usepackage{comment}
\usepackage[usenames,dvipsnames]{xcolor}
\usepackage{empheq}
\usepackage{systeme}

\DeclareMathOperator{\sgn}{sgn}
\usepackage{amsmath}

\begin{document}

\begin{titlepage}
\renewcommand{\thefootnote}{\fnsymbol{footnote}}
\vspace*{-2cm}
\begin{flushright}
\jobname .pdf\\ 
\today \\

\end{flushright}

\vspace*{1cm}

\begin{center}
{\huge {\bf Integrability and cycles of deformed $\CN=2$ gauge theory }}

\vspace{10mm}
{\Large Davide Fioravanti, Daniele Gregori}
\\[.6cm]
{\em {}Sezione INFN di Bologna\\ 
Dipartimento di Fisica e Astronomia,
Universit\`a di Bologna} \\
{\em Via Irnerio 46, 40126 Bologna, Italy}\\
[.4cm]
\texttt{ fioravanti\,@\,bo.infn.it ,\quad dagregor\,@\,bo.infn.it}
\end{center}

\vspace{0.7cm}

\begin{abstract}
\noindent

To analyse pure $\CN=2$ $SU(2)$ gauge theory in the Nekrasov-Shatashvili (NS) limit (or deformed Seiberg-Witten (SW)), we use the Ordinary Differential Equation/Integrable Model (ODE/IM) correspondence, and in particular its (broken) discrete symmetry in its extended version with {\it two} singular irregular points. Actually, this symmetry appears to be 'manifestation' of the spontaneously broken $\mathbb{Z}_2$ R-symmetry of the original gauge problem and the two deformed SW cycles are simply connected to the Baxter's $T$ and $Q$ functions, respectively, of the Liouville conformal field theory at the self-dual point. The liaison is realised via a second order differential operator which is essentially the 'quantum' version of the square of the SW differential. Moreover, the constraints imposed by the broken $\mathbb{Z}_2$ R-symmetry acting on the moduli space (Bilal-Ferrari equations) seem to have their quantum counterpart in the $TQ$ and the $T$ periodicity relations, and integrability yields also a useful Thermodynamic Bethe Ansatz (TBA) for the cycles ($Y(\theta,\pm u)$ or their square roots, $Q(\theta,\pm u)$). A latere, two efficient asymptotic expansion techniques are presented. Clearly, the whole construction is extendable to gauge theories with matter and/or higher rank groups.

\vspace{0.5cm}
\end{abstract}

\vfill

\end{titlepage}
\vfil\eject

\setcounter{footnote}{0}

%
%
%

\section{Essentials of $\CN=2$ $SU(2)$ Seiberg-Witten gauge theory}
\label{essentials}

According to Seiberg-Witten theory~\cite{SeibergWitten1994}, the low energy effective Lagrangian of 4d $\CN =2$ SUSY $SU(2)$ pure gauge theory is expressed through an holomorphic function $\CF_{\text{SW}}(a^{(0)})$ called \emph{prepotential}. It may be thought of as constructed from the Seiberg-Witten cycle $a^{(0)} =2 \langle \Phi \rangle$ ($\Phi$ is the scalar field) and its \emph{(Legendre) dual $a^{(0)}_D =\partial \CF_{\text{SW}}/ \partial a^{(0)}$}: 
\begin{align}
 a^{(0)}(u, \Lambda)
&=\frac{1}{2 \pi} \int_{-\pi}^{\pi} \sqrt{2u-2\Lambda^2 \cos{z} } \, dz =\Lambda\sqrt{2(u/\Lambda^2+1)}\,\, {}_2F_1(-\frac{1}{2},\frac{1}{2},1;\frac{2}{1+u/\Lambda^2}) \,\, , \label{a0}  \\
 a^{(0)}_D(u, \Lambda)
&= \frac{1}{2\pi} \int_{- \arccos(u/\Lambda^2)-i0}^{ \arccos(u/\Lambda^2)-i0} \sqrt{2u-2\Lambda^2 \cos{z} } \, dz    =  - i \Lambda \frac{(u/\Lambda^2-1)}{2} \,\, {}_2F_1(\frac{1}{2},\frac{1}{2},2;\frac{1-u/\Lambda^2}{2})  \, , \label{a0D} 
\end{align}
which are functions of the modulus $u =\langle \text{tr} \, \Phi^2 \rangle $ (for fixed parameter $\Lambda$\footnote{We may calculate the first integral for $u>\Lambda^2$ while the second one for $u<\Lambda^2$ along a {\it continuous} (without jumps, and hence changing sheet) path in $z$ and then analytically continue in $u$; we will analyse better the complex structure below, in Section \ref{TandQ}.}) upon eliminating $u$ to obtain $a^{(0)}_D( a^{(0)})$ (and finally integrating). The $\CN =2$ SYM classical action enjoys a $U(1)_{\mathcal{R}}$ $\mathcal{R}$-symmetry, which is broken to $\mathbb{Z}_8$ by one-loop and instanton corrections. Eventually it is broken down to $\mathbb{Z}_4$ by the vacuum, so that the (spontaneously) broken part, which is a $\mathbb{Z}_2$, {\it i.e.} $u \rightarrow -u$, connects two equivalent vacua \cite{SeibergWitten1994}: we will see that somehow {\it this} broken symmetry plays an important r\^ole also in the deformed theory.

The exact partition function for $\CN =2$ SYM theories, with all instanton corrections, has been obtained through equivariant localisation techniques in~\cite{Nekrasov2003, NekrasovOkounkov2006}: two super-gravity parameters, $\epsilon_1$ and $\epsilon_2$,  \emph{the omega background} deform space-time. When both $\epsilon_1\, ,\epsilon_2\rightarrow 0$, the logarithm of the partition function reproduces the Seiberg-Witten prepotential $\CF_{\text{SW}}$~\cite{NekrasovOkounkov2006}. The latter can also be thought of as a successive limit of the Nekrasov-Shatashvili (NS) limiting theory \cite{Nekrasov2009}, defined by the quantisation/deformation (of SW) $\epsilon_1= \hslash$, $\epsilon_2 \to 0$.

More specifically, having in mind the AGT corresponding Liouville field theory \cite{AldayGaiottoTachikawa:2010,Gaiotto:2009} and precisely its level $2$ degenerate field equation \cite{AwataYamada:2009}, we may think of it as a quantisation/deformation\footnote{We shall prefer this latter connotation as the former generates sometimes confusion with gauge theory quantisation.} of the {\it square} of the SW differential which takes up the form of the Mathieu equation
\begin{equation}
-\frac{\hslash^2}{2} \frac{d^2}{dz^2} \psi(z) +[ \Lambda^2 \cos{z}-u ] \psi(z)= 0\, . \label{SWeq}
\end{equation} 
For the deformed prepotential $\CF_{\text{NS}}$ (logarithm of the partition function)  may be derived as above by eliminating $u$  between the two deformed cycles \cite{MironovMorozov:2009}
\begin{align}
a (\hslash,u, \Lambda)
&=\frac{1}{2\pi}\int_{-\pi}^{\pi} \mathcal{P}(z; \hslash,u, \Lambda) \, dz \,\,\, , \qquad
 a_D(\hslash,u, \Lambda)
=\frac{1}{2\pi}  \int_{- \arccos{(u/\Lambda^2)-i0}}^{ \arccos{(u/\Lambda^2)-i0}} \mathcal{P}(z;\hslash, u, \Lambda)\, dz    \,    \label{NScycles} 
\end{align}
(in gauge theory $a =2 \langle \tilde \Phi \rangle $) of the quantum SW differential $\mathcal{P}(z) =- i \frac{d}{dz} \ln{\psi}(z)$. In particular, we may expand asymptotically, around $\hslash = 0$, $ \mathcal{P} (z) \doteq \sum_{n=-1}^\infty \hslash^n  \mathcal{P}_n(z)$,  and then the NS-deformed cycles (modes) are 
\begin{align}
  a^{(n)}(u, \Lambda)
&= \frac{1}{2\pi}\int_{-\pi}^{\pi} \mathcal{P}_{2n-1}(z; u, \Lambda) \, dz \qquad
 a^{(n)}_D(u, \Lambda)
= \frac{1}{2\pi} \int_{- \arccos{(u/\Lambda^2)}-i0}^{ \arccos{(u/\Lambda^2)}-i0} \mathcal{P}_{2n-1} (z; u, \Lambda)\, dz    \, .   \label{ananD} 
\end{align}
Alternatively, we can use Matone's formula for the prepotential \cite{Matone:1995}, generalised for the deformations in \cite{FlumeFucitoMoralesPoghossian:2004}.

This letter is organised as follows. In Section~\ref{NSdefComp} we develop a very efficient and general idea of computing the large energy and small $\hslash$ WKB expansion of the wave function ({\it cf.} also ~\cite{FachechiFioravanti}), which we apply to give efficient recursive formul\ae \, for the NS-deformed cycles modes. In Section~\ref{Zam} we present the analysis of the ODE/IM correspondence for the Liouville integrable model, deeply based on an unfinished work~\cite{ZamolodchikovMemorial} by the late scholar Al. B. Zamolodchikov. In Section ~\ref{LIM} we apply the results of Section~\ref{NSdefComp} to efficiently compute the general Liouville local integrals of motion. In Section~\ref{TandQ} we interpret the Baxter's $T$ and $Q$ functions as the NS-deformed Seiberg-Witten cycles. In Section~\ref{functional} we interpret the $QQ$ relation as a peculiar gauge TBA and the $TQ$ relation with the $T$ periodicity relation as quantum $\mathbb{Z}_2$ symmetry relations. In the last Section \label{conclusions-matter} we comment on R-symmetry, (chiral) gauge theories with matter and other groups and some perspectives.

\section{One-step large energy/WKB recursion and deformed cycles modes} \label{NSdefComp}

This Section contains, as a preamble, a general two-fold result concerning the (modified) Schr\"odinger equation: an efficient technique of one-step-recursion for computing the asymptotic expansion of the wave function/cycles, for high energy and small Planck constant. In the second part of this Section, we will apply this result for efficiently compute the NS-deformed cycles modes~\eqref{ananD}.

Let us first start by the large energy expansion of the wave function which we will apply for computing the local integrals of motion for Liouville theory in section \ref{LIM}. Consider a general \emph{modified Schr\"odinger equation}, with {\it energy} $e^{2 \theta}$ which multiplies the modification $\phi(x)$ and potential $v(x)$ 
\begin{equation}
\biggl\{-\frac{d^2}{d x^{2}} + v(x)- e^{2 \theta} \phi(x)   \biggl\}\psi(x) = 0 \label{mSchEquation}
\end{equation}
By the transformation $dw = \sqrt{\phi} dx$, $\chi =\sqrt[4]{\phi}\psi$, the modified Schrödinger equation can be transformed into an ordinary Schrödinger equation 
 \be
\biggl\{-\frac{d^2}{d w^{2}} + U(w)- e^{2 \theta}   \biggl\}\chi(w) = 0, \,\,\,\,\,\,\,\, U = \frac{v}{\phi}+\frac{1}{4}\frac{\phi''}{\phi^2} -\frac{5}{16} \frac{\phi^{'2}}{\phi^3} \,\,\, . \label{Sch-eq}
\ee 
As usual, we define $\Pi(w)= - i \frac{d}{dw} \ln{\chi}(w)\doteq \sum_{n=-1}^\infty e^{-n\theta}  \Pi_n(x)$ (the last equality is {\it asymptotical} for large energy, $\Re \theta\rightarrow +\infty$) satisfying the usual Riccati equation
\be
\Pi^2(w) - i \frac{d\Pi(w)}{dw} = e^{2 \theta}-U(w)
\ee
which is solved by $\Pi_{-1}= 1$, $\Pi_0= 0$, $\Pi_1= - \frac{1}{2}U$\footnote{It is solved also by the other solution generated by $\Pi_{-1}= -1$, $\Pi_0= 0$, $\Pi_1= \frac{1}{2}U$.} and the recursion relation for the high energy modes 
\be
 \Pi_{n+1} =+ \frac{1}{2} \left \{i \frac{1}{\sqrt{\phi(x)}}\frac{d\Pi_n}{dx} - \sum_{m=1}^{n-1} \Pi_m \Pi_{n-m}\right \} \qquad n =1,2,...\label{Srecursive} \,\, .
\ee
Eventually the wave function $\psi=(\phi)^{-1/4} \chi$ can be written and then expanded at large energy $\Re \theta\rightarrow +\infty$
\begin{equation}
\psi(x; \theta)=  \frac{1}{\sqrt[4]{\phi(x)}}  
 \exp \Bigl \{  i \int^x \sqrt{\phi(x')} \, \Pi(x') dx' \Bigr \} \doteq \frac{1}{\sqrt[4]{\phi(x)}} \exp \Bigl \{ i \sum_{n=-1}^\infty e^{-n\theta} \int^x \sqrt{\phi(x')} \, \Pi_n(x') dx' \Bigr \} \, ,
\label{P} 
\end{equation}
from which we can read off the quantum momentum of the (modified) Schr\"odinger equation \eqref{mSchEquation}: 
\be
\CP(x)= - i \frac{d}{dx} \ln{\psi}(x) = \frac{i}{4} \frac{\phi'}{\phi}+ \sqrt{\phi(x)} \, \Pi(x) \doteq \sum_{n=-1}^\infty e^{- n \theta}  \mathcal{P}_n(x) \,\, .
\label{P-Pi}
\ee
Then, we split $\Pi(x)= \Pi_{odd}(x)+\Pi_{even}(x)$ into odd and even indexes and notice immediately that $\Pi_{even} = -\frac{1}{2} (\ln{ \Pi_{odd}})'$ is a total derivative, which becomes irrelevant when integrating under specific circumstances, for example on the real axis with suitable asymptotic conditions or on a period. Forgetting about the even modes $\Pi_{2n}$ (total derivatives), an important fact happens for the large energy expansion of \eqref{Sch-eq}, {\it i.e.} the arising of the Gelfand-Dikii (differential) polynomials, $R_n[U]$~\cite{GelfandDikii:1975}. They satisfy this \emph{one-step} recursion relation
\begin{align}
\begin{split}
\frac{dR_{n+1}}{dx} &= - \frac{1}{4} \frac{1}{\phi} \frac{d^3}{dx^3} R_n + \frac{3}{8} \frac{\phi'}{\phi^2} \frac{d^2}{dx^2} R_n + \Bigl[ \frac{v}{\phi} + \frac{3}{8} \frac{\phi''}{\phi^2} - \frac{9}{16} \frac{\phi^{'2}}{\phi^3} \Bigr] \frac{d}{dx} R_n \\
&+ \Bigl[ \frac{1}{2}\frac{v'}{\phi} -\frac{1}{2} \frac{v \phi'}{\phi^2} + \frac{1}{8} \frac{\phi'''}{\phi^2} - \frac{9}{16} \frac{\phi'' \phi'}{\phi^3} + \frac{15}{32} \frac{\phi^{'3}}{\phi^4} \Bigr] R_n 
\label{GDrecE}
\end{split}
\end{align}
with initial condition $R_0=1$. In fact, it can be proven (see for instance \cite{FachechiFioravanti} for a detailed discussion) that they are proportional to the modes $\sqrt{\phi(x)}\Pi_{2n-1}$ up to a $x$-total derivative:
\begin{equation} 
\Pi_{2n-1}(x) = \frac{-1}{2n-1}  R_n(x) + \frac{d}{dw} ( \text{local fields}) .
\label{energyequiv} 
\end{equation}

The advantage of using the equivalent $R_n$ integrands (which are \emph{equivalent} under integration, if one can neglect total derivatives) is that their recursion~\eqref{GDrecE} is far simpler than that~\eqref{Srecursive} for the original integrands $\mathcal{P}_{2n-1}$. In fact, in~\eqref{GDrecE}, to compute the $n+1$-th term, it is sufficient to know \emph{only the first preceding} $n$-th term, \emph{not all} the preceding, as in~\eqref{Srecursive}.

We show now that these results can be adapted for the usual small $\hslash$ WKB asymptotic expansion of a generic Schrödinger equation
\begin{equation}
+ \frac{d^2}{dx^2} \psi +\frac{\phi(x)}{\hslash^2} \psi = 0 \qquad \text{with} \quad \phi(x) =2m (E- V(x)  ) \, .
\label{Schr}
\end{equation} 
In fact, the usual WKB analysis envisages the exact quantum momentum $\mathcal{P}(x)= - i \frac{d}{dx} \ln{\psi}(x) \doteq \sum_{n=-1}^\infty \hslash^n  \mathcal{P}_n(x)$ verifying the Riccati equation and modes recursion relation, respectively
\begin{equation}
\mathcal{P}^2(x) - i \frac{d\mathcal{P}(x)}{dx} = \frac{\phi(x)}{\hslash^2}
\,\, ,\,\,\,\, \mathcal{P}_{n+1} = \frac{1}{2\sqrt{\phi} } \left(i \frac{d}{dx} \mathcal{P}_n- \sum_{m=0}^{n} \mathcal{P}_m \mathcal{P}_{n-m} \right)\,\, , \label{WKBStandardRec}
\end{equation}
with initial condition the classical momentum $\mathcal{P}_{-1} = \sqrt{\phi}$ \footnote{As above there is also the solution with $\mathcal{P}_{-1} =-\sqrt{\phi}$.}. As above, we split $\mathcal{P}(x)= \mathcal{P}_{odd}(x)+\mathcal{P}_{even}(x)$: then $\mathcal{P}_{even} = -\frac{1}{2} (\ln{ \mathcal{P}_{odd}})'$ and $\mathcal{P}_{2n}$ are total derivatives, which, under specific circumstances, can be forgetten about. Now we wish to think of \eqref{Schr} as the particular case, $v=0$, of the previous modified Schr\"odinger equation \eqref{mSchEquation} with {\it energy} $=1/\hslash^2$. Thus, we obtain the usual Schr\"odinger equation \eqref{Sch-eq} with potential $U=\frac{1}{4}\frac{\phi''}{\phi^2} - \frac{5}{16}\frac{\phi^{'2}}{\phi^3}$: in this manner small $\hslash$ is interpreted as large energy. And we can make us of the Gelfand-Dikii polynomials~\cite{GelfandDikii:1975}, with recursion relation \eqref{GDrecE} with $v=0$
\begin{align} 
&R_{n+1}' = - \frac{1}{4\phi} R_n''' + \frac{3}{8} \frac{\phi'}{\phi^2} R_n'' + \Bigl(\frac{3}{8} \frac{\phi''}{\phi^2} -\frac{9}{16} \frac{\phi^{'2}}{\phi^3} \Bigr) R_n'  + \Bigl( \frac{1}{8} \frac{\phi'''}{\phi^2} -\frac{9}{16} \frac{\phi'' \phi'}{\phi^3} +\frac{15}{32} \frac{\phi^{'3}}{\phi^4} \Bigr) R_n \, ,  \label{Trecursive}
\end{align} 
and initial condition $R_0= 1$. In fact, we have seen above $\mathcal{P}_{2n-1}= \sqrt{\phi(x)}\Pi_{2n-1}$ which, in its turn, is expressible by $\sqrt{\phi(x)}R_n$ up to a $x$-total derivative:
\begin{equation}
\mathcal{P}_{2n-1}(x) = \frac{ -1 }{2n-1}\sqrt{\phi(x)} R_n(x) + \frac{d}{dx} ( \text{local fields}) \,\, .\label{equiv_h}
\end{equation}
The advantage of using the $R_n$ integrands (which are {\it equivalent} as they give the same integral, under suitable boundary conditions) is that their recursion~\eqref{Trecursive} is far simpler than that~\eqref{WKBStandardRec} for the original integrands $\mathcal{P}_{2n-1}$. In fact, using recusion~\eqref{Trecursive} to compute the $n+1$-th term, requires to know \emph{only the first preceding} $n$-th term, \emph{not all} the preceding, as in recursion~\eqref{WKBStandardRec}. Now, we can efficiently calculate the Nekrasov-Shatashvili deformed integrals. The equation to be considered is the Mathieu equation~\eqref{SWeq}, for which $\phi(z) = 2\Lambda^2 \cos{z} -2u$. The Gelfand-Dikii polynomials can be expanded in the basis of the inverse powers of $\phi(z)^{-m}$, $n \leq m \leq 3n$, with some coefficients which satisfy a one-step recursion relation. Eventually, the cycle are given linear differential operators acting on the original SW cycles:
\begin{align}
\begin{pmatrix}
 a^{(n)}(u,\Lambda) \\
  a^{(n)}_D(u,\Lambda) 
  \end{pmatrix}&= \sum_{m=n}^{3n} \alpha_{n,m}(u,\Lambda) \frac{\partial^{m}}{\partial u^{m}}  \begin{pmatrix}
 a^{(0)}(u,\Lambda) \\
  a^{(0)}_D(u,\Lambda) 
  \end{pmatrix} \quad 	 \label{operatorB_nm} 
\end{align}
with coefficients given by this one-step recursion
\begin{align}
\alpha_{n+1,m+1} &=- \frac{n-\frac{1}{2}}{n+\frac{1}{2}}\left \{ \frac{ \frac{1}{8}(m+\frac{1}{2})^3}{(m+1)(m-\frac{1}{2})} \alpha_{n,m} +\frac{\frac{1}{4}  m(m+\frac{1}{2})u}{(m+1)(m-\frac{3}{2})} \alpha_{n,m-1} +\frac{\frac{1}{8} (m+\frac{1}{2})(u^2-\Lambda^4)}{(m+1)(m-\frac{5}{2})} \alpha_{n,m-2}\right \} \label{recB_nm}
\end{align}
and initial condition $\alpha_{0,0}= 1$. This relation is a very efficient way of computing the NS-deformed cycle modes. From these coefficient we can also derive those in the linear differential operators of \cite{He-Miao} (and prove thus their conjecture). By commuting the latter with the Picard-Fuchs equation for SW \cite{BilalFerrari:1996}
\be
\left \{ (u^2-\Lambda^4) \frac{\partial^2}{\partial u^2} + \frac{1}{4} \right \} a^{(0)}_D(u,\Lambda) =0\, , 
\label{PF0}
\ee
we gain straightforwardly an explicit formula for computing the coefficients of all the {\it quantum Picard-Fuchs equations} (constraining both cycles $a^{(n)}_D(u,\Lambda)$ and $a^{(n)}(u,\Lambda)$), {\it e.g.}:
\begin{align}
\left \{ (u^2-\Lambda^4) \frac{\partial^2}{\partial u^2} + 4 u \frac{\partial }{\partial u}+ \frac{5}{4} \right \} a^{(1)}_D(u,\Lambda) &=0 \label{PF1} \\
\left[(u^2-\Lambda^4)\frac{\partial^2}{\partial u^2 } + 6 u\frac{\frac{u^2}{\Lambda^4}+\frac{111}{8}}{ \frac{u^2}{\Lambda^4}+\frac{325}{32}}  \frac{\partial }{\partial u} + \frac{21}{4}\frac{\frac{u^2}{\Lambda^4}+\frac{689}{32}}{ \frac{u^2}{\Lambda^4}+\frac{325}{32}}\right ] a^{(2)}_D(u,\Lambda) &= 0 \label{PF2}\, .
\end{align}
Eventually, from the knowledge of the cycles we can determine the partition function by different means as explained in Section \ref{essentials}.

\section{Liouville ODE/IM} \label{Zam}

The original ODE/IM correspondence establishes an exact parallel between a particular Schr\"odinger equation (ODE) and the Minimal Conformal models (IM) ~\cite{DoreyTateo1998,BazhanovLukyanovZamolodchikov2001}, without masses ({\it cf.} also ~\cite{Suzuki1999}). Later, it was extended to the massive case ~\cite{LukyanovZamolodchikov:2010}.  Although there was already a bold suggestion already in ~\cite{DoreyTateo1999}, the conjecture for the (conformal) Liouville field theory came only in a brilliant draft paper ~\cite{ZamolodchikovMemorial} by the late scholar Al. B. Zamolodchikov and takes the form of the \emph{Generalized Mathieu equation} (GME): 
\begin{equation} \label{GMB} 
\biggl\{-\frac{d^2}{d y^2} + e^{\frac{\alpha + y}{b}}+e^{b(\alpha - y)} +P^2 \biggl\}\psi(y) = 0 \,\, .
\end{equation}
The parameters $b$ and $P$ are respectively the \emph{Liouville coupling} and \emph{momentum} and express the central charge $c=~1+6(b +b^{-1})^2$ and the conformal weight $\Delta=~(c-1)/24-P^2$. Liouville field theory enjoys a duality symmetry for $b \to 1/b$ (\emph{self-dual} point $b=1$). Following  ~\cite{DoreyTateo1999}, we may immagine that this equation could be obtained \emph{heuristically} from the ODE/IM equation for the minimal models, through some continuation in $\beta= i b$ and some transformation on the independent variable (crucially, the Langer transform, $x=e^y$). However, we found the form \eqref{GMB} not adequate for the large rapidity expansion, as $e^\alpha$ appears with two different powers. We have solved this problem by the shift $y \to y +  \alpha \frac{b-1/b}{b+1/b}$, after which the GME acquires the modified Schrödinger form: 
\begin{equation} \label{GMF} 
\biggl\{-\frac{d^2}{d y^2} + e^{2\theta}(e^{y/b}+e^{- yb}) +P^2 \biggl\}\psi(y) = 0 
\end{equation}
with the rapidity $\theta$ defined as $\theta = \alpha /(b +b^{-1})$. In the rest of this Section we will summarise our understanding of draft paper ~\cite{ZamolodchikovMemorial} by using the GME~\eqref{GMF}. It has the subdominant asymptotic solutions: for $\Re{y} \to+ \infty$, within $ | \Im{( \theta +\frac{y}{2b})}| < \frac{3}{2} \pi $ and for $\Re{y} \to - \infty$, within $| \Im{( \theta- \frac{by}{2})}| < \frac{3}{2} \pi$, respectively
\begin{align}
U_0(y) &\simeq \frac{1}{\sqrt{2}} \exp \Bigl \{ - \theta/2 - y/4b \Bigr \} \exp \biggl \{  - 2b e^{\theta + y/2b} \biggr \} \qquad \Re{y} \to+ \infty \, \, ; \label{U0lead}\\ 
V_0(y) &\simeq \frac{1}{\sqrt{2}} \exp \Bigl \{ - \theta/2 + yb/4 \Bigr \} \exp \biggl \{  -  \frac{2}{b}  e^{\theta - yb/2}\biggr \} \qquad \Re{y} \to - \infty\, \, . \label{V0lead}
\end{align}
Other solutions can be generated applying on these the following discrete symmetries of the GME~\eqref{GMF}
\begin{align}
\Lambda_b &: \theta \to \theta + i \pi \frac{b}{q}  \qquad y \to y +\frac{2  \pi i }{q}  \quad , \quad \Omega_{b} : \theta \to \theta + i \pi \frac{1}{bq} \qquad y \to y  -\frac{2  \pi i }{q} 
\label{discr-symm}
\end{align}
where $q = b + 1/b$: concisely $U_k =\Lambda^k_b U_0$ and $V_k=\Omega^k_{b} V_0$, with $U_k$ invariant under $\Omega_{b}$ and $V_k$ under $\Lambda_b$. We may interpret this phenomenon as a spontaneous symmetry breaking for the differential equation (vacua are the solutions). Now we apply these (broken) symmetries to derive interesting functional and integral equations for the gauge theory. On the other hand, the symmetry $\Pi : \theta\rightarrow \theta+i\pi$ would not do the same job in the present case with two irregular singularities as it transforms {\it simultaneously}   $U_0 \rightarrow U_1$ and $V_0\rightarrow V_1$ (differently from \cite{GaiottoMooreNeitzke:2008} and \cite{marino2018} with only one irregular singularity, see also \cite{FR-ampilett} for a detailed examination of the two kinds of symmetries).

In fact, we will prove correct (as conjectured by \cite{ZamolodchikovMemorial}) to define the Baxter's $Q$ function as the wronskian 
\begin{equation} 
Q(\theta, P^2) = W[U_0,V_0] \label{QW} \, \, .
\end{equation}
Definition \eqref{QW} gives rise to $Q(\theta + i \pi p) =W[U_1,V_0] (\theta)$ upon action of $\Lambda_b$: these are equivalent to the linear dependence 
\be
i V_0(y) = Q(\theta + i \pi p )U_0(y) - Q(\theta )U_1(y) \, ,
\label{lin0}
\ee
where $p=b/q$ (from the asymptotic calculation $W[U_1,U_0]= i$). Which is transformed by $\Omega_{b}$ into 
\be
 i V_1(y) = Q(\theta + i \pi ) U_0(y) -Q(\theta + i \pi (1-p))U_1(y)  \, ,
\label{lin1}
\ee
namely $Q(\theta +  i \pi (1-p) )=W[U_0,V_{1}](\theta)$ and $Q(\theta +  i \pi  )=W[U_1,V_{1}](\theta)$. The basilar functional relation (anticipated for the massive theory by other means in \cite{Zamolodchikov:2000}), the $QQ$ relation is obtained by taking the wronskian $W[V_0,V_1]$ ($=i$ from asymptotics) between the right hand sides
\begin{equation}
 1+ Q(\theta + i \pi (1-p))Q(\theta+  i \pi p)= Q(\theta + i \pi ) Q(\theta )  \, . \label{QQZ}
\end{equation}
If one defines the two (dual) $T$ functions as 
\begin{equation}
T(\theta) = Q(\theta - i \pi p) Q(\theta + i \pi ) - Q(\theta + i \pi p ) Q(\theta + i \pi (1-2p) ) \, , \quad \tilde{T}(\theta) = T(\theta) \Bigr |_{b \to 1/b}\, , \label{TZ}
\end{equation}
(also $T=-i\, W[U_{-1},U_1]$ and $\tilde{T}=i \, W[V_{-1},V_1]$) by using the $QQ$ relation~\eqref{QQZ}, these two Baxter's $TQ$ relations follow
\begin{equation}
T(\theta) Q(\theta) = Q(\theta + i \pi p ) + Q(\theta - i \pi p ) \qquad \tilde{T}(\theta) Q(\theta) = Q(\theta + i \pi (1-p)) + Q(\theta - i \pi (1-p))\, , \label{TQZ}
\end{equation}
as well as the periodicity of $T$ \cite{ZamolodchikovMemorial}
\begin{equation}
T(\theta + i \pi (1-p) ) = T(\theta) \qquad \tilde{T}(\theta + i \pi p ) = \tilde{T}(\theta) \, \, . \label{TperZ}
\end{equation}
Also the Liouville $Y$-system can be obtained from the $QQ$-system, by defining $Y(\theta) =Q(\theta + i \pi a/2 )Q(\theta-  i \pi a/2 )$, where $a = 1-2p$ 
\begin{equation} 
Y(\theta + i \pi /2)Y(\theta - i \pi /2) = \Bigl (1+Y(\theta + i a \pi /2)\Bigr)\Bigl(1 + Y(\theta - i a \pi /2)\Bigr) \label{Ysystem} .
\end{equation} 
This functional equation can be inverted into the Thermodynamic Bethe Ansatz (TBA) for the pseudoenergy $\varepsilon(\theta) = - \ln Y(\theta)$  
\begin{align}
\varepsilon(\theta)   &=  \frac{8\sqrt{\pi^3}\,q}{\Gamma(\frac{b}{2q} )\Gamma(\frac{1}{2bq})} e^{\theta} -  \int_{-\infty}^\infty  \left [ \frac{1}{\cosh (\theta-\theta' + i a \pi /2) } +\frac{1}{\cosh (\theta-\theta' - i a \pi /2) }  \right ] \ln \left [ 1 +  \exp \{ - \varepsilon(\theta') \} \right ] \frac{d \theta'}{ 2 \pi} \, ,
\label{LiouvilleTBA}
\end{align} 
with the coefficient of the forcing term (zero-mode) fixed as given by the leading order of $Q$ below, \eqref{QexpI2n-1}. This TBA equation goes into that in  ~\cite{ZamolodchikovMemorial, Zamolodchikov:2000, Lukyanov2001} upon a real shift on $\theta$. In it $P$ does not appear explicitly, but (numerically) in the asymptotic linear behaviour of $\varepsilon(\theta, P^2) \simeq +4 q P \theta$, $P>0$, at $\theta \to - \infty$ \cite{ZamolodchikovMemorial}, which matches the analytic computation of the wronskian (\ref{QW}) via $1+Y(\theta) =Q(\theta + i \pi/2 )Q(\theta-  i \pi/2 )$ (on the other hand from \eqref{Ysystem} we only know that $Y$ must diverge).  

\noindent
The self-dual GME ($b=1$ in \eqref{GMF}) is known in literature as {\it modified Mathieu equation}:
\begin{equation} \label{GMF1} 
\biggl\{-\frac{d^2}{d y^2} +2 e^{2\theta} \cosh y +P^2 \biggl\}\psi(y) = 0 \,\, , 
\end{equation} 
and is the non-compact version of equation \eqref{SWeq}, so establishing a contact with gauge theory (which importantly exhibits two irregular singularities). In particular, the discrete symmetry (\ref{discr-symm}) is an enhanced (by the covering  $y=\ln x$) version of the original $\mathbb{Z}_2$ spontaneously broken symmetry (in the $x$ variable) of SW \cite{SeibergWitten1994}. Since $a= 0$ then $Q^2 =Y= \exp [ - \varepsilon ]$ and the TBA becomes an integral equation for the Baxter's $Q$ function~\cite{ZamolodchikovMemorial}
\begin{align}
\ln{Q(\theta) }  &=- \frac{8\sqrt{\pi^3}}{\Gamma^2(\frac{1}{4} )} e^{\theta}   +\int_{-\infty}^\infty \frac{ \ln \left [ 1 +Q^2(\theta') \right ]}{\cosh{(\theta- \theta')}} \frac{d \theta'}{ 2 \pi} \, .
\end{align}

\section{One-step large energy recursion and local integrals of motion} \label{LIM}

We wish here to compute the Baxter's $Q$ function $Q$ and then the Liouville Local Integrals of Motion (LIM). About $Q$, ~\eqref{lin0} says that it can regarded as the regularised value of the solution $V_0$ \eqref{V0lead} at $y \to + \infty$:
\begin{equation}
Q(\theta) =-i \lim_{y \to +\infty} \frac{V_0(y; \theta)}{U_1(y; \theta)} = \sqrt{2} e^{\frac{\theta}{2}} \lim_{y \to +\infty} e^{\frac{y}{4b} -2b e^{\theta+\frac{y}{2b}}}V_0(y; \theta) \,\, .
\label{Qdef}
\end{equation}
We can write $V_0$~\eqref{V0lead} in terms of $\Pi (w)=-i \, d\ln(\sqrt[4]{c_b(y)} V_0(w))/d w$ in a convergent form of (\ref{P}) 
\begin{align}
V_0(y; \theta)&= \frac{e^{-\frac{\theta}{2}}}{\sqrt{2} \sqrt[4]{c_b(y)} }  \exp \biggl \{  - \frac{2}{b}e^{\theta - \frac{by}{2}}+ 2b e^{\theta +\frac{y}{2b}}+\int_{-\infty}^y  \Bigl [   \sqrt{c_b(y')} \Pi(y'; \theta) - e^{\theta}(e^{-\frac{by'}{2}} +e^{\frac{y'}{2b}}) \Bigr] dy' \biggr \}  \label{V0S}
\end{align}
where $c_b(y) =  - \phi(y) =e^{y/b}+e^{- yb}$  and $d w=\sqrt{\phi(y)} dy = -i \sqrt{c_b(y)} dy$. Now, we can write the asymptotic expansion of the $Q$ function~\eqref{Qdef} for $\theta \to + \infty$, by using formula~\eqref{energyequiv}, as we are integrating on $\mathbb{R}$ and hence can neglect the decaying derivatives \be
\ln Q(\theta )=\int_{-\infty}^{+\infty}  \Bigl [   \sqrt{c_b(y)} \Pi(y) - e^{\theta}(e^{-\frac{by}{2}} +e^{\frac{y}{2b}}) \Bigr] dy  \doteq    e^{\theta}  \int_{-\infty}^\infty \left[  \sqrt{c_b(y)}-e^{\frac{y}{2b}} -e^{-\frac{by}{2}} \right]dy  -\sum_{n=1}^\infty \frac{e^{\theta(1-2n)}}{2n-1} \int_{-\infty}^\infty  \sqrt{c_b(y)} R_{n}(y)\,dy  \,\, . \label{Xexp}
\ee  
Notice for the future developments that $\ln Q$ is given $i$ times the integral of the regularised momentum
\be
\CP_{reg}(y)=-i \sqrt{c_b(y)} \Pi(y) + i e^{\theta}(e^{-\frac{by}{2}} +e^{\frac{y}{2b}})=\CP(y) + i e^{\theta}(e^{-\frac{by}{2}} +e^{\frac{y}{2b}}) - \frac{i}{4} \frac{c_b'}{c_b}
\label{Preg}
\ee 
thanks to \eqref{P-Pi}: this fact is valid for any $b$ and connects $Q$ to SW-NS cycles ({\it cf.} below the development for the pure gauge case $b=1$). Moreover, upon identification of the $n$-th local integral of motion $I_{2n-1}$ up to an arbitrary normalisation $C_n$
\be
\label{IR}
C_n I_{2n-1} =  \frac{1 }{2n-1} \int_{-\infty}^\infty \sqrt{c_b(y)} R_n(y)\, dy\,\, , 
\ee
they are given by the large $\theta$ asymptotic expansion of the Baxter's $Q$ function \eqref{Xexp}: 
\begin{equation}
Q(\theta, P^2 , b) \doteq    \exp  \left \{-e^{\theta} \frac{4\sqrt{\pi^3}q}{ \sin{(\pi b/q)}\Gamma(\frac{b}{2q}) \Gamma(\frac{1}{2bq})}   -\sum_{n=1}^\infty e^{\theta(1-2n)} C_n(b) I_{2n-1}(b,P^2)\right \} \,\, .\label{QexpI2n-1}
\end{equation}
For instance, we can use the normalisation constants: 
\begin{equation}
C_n(b) = \frac{\Gamma(\frac{(2n-1)b}{2q}  ) \Gamma(\frac{(2n-1)}{2bq})}{2 \sqrt{\pi} n! q}  \label{BnL} \,\, .
\end{equation}
This expansion matches the numerical results from TBA \eqref{LiouvilleTBA}. Now, we can make explicit the one-step recusion procedure \eqref{GDrecE} for the $R_n$ in this particular case ~\eqref{GMF}. In fact, we can expand the $R_n$ in a certain basis of functions whose integrals turn out to be the Euler Beta function and whose coefficients $\lambda_{n,m}$, with $n \leq m \leq 3n$ satisfy a one-step recursion. We will give the details elsewhere and just give the final formula for the LIMs
\begin{align}
I_{2n-1}(b, P^2) &= \frac{(-1)^{n}(2n)!!}{(2n-1)!!} \sum_{m=n}^{3n}  \frac{\Gamma(n-1/2)}{\Gamma(n-1/2+m-n)}  \frac{\Gamma(\frac{n-1/2}{bq} +m-n)}{\Gamma(\frac{n-1/2}{bq})}                  \lambda_{n,m}(b,P^2) \,\, , \label{I2n-1}
\end{align}
with the recursion for the coefficients
\begin{align}
\lambda_{n+1,m+1} &= \sum_{k=m+1}^{3n+3} \frac{m!}{k!q}  \frac{\Gamma(-\frac{(n+1)b}{q} +k+1)}{\Gamma(-\frac{(n+1)b}{q} +m +2)}  \sum_{l=0}^{3}  F_l(n,k-l)\,\lambda_{n,k-l} \,\, ,\label{GDrecFinalLiouville} 
\end{align}
from the initial condition $\lambda_{0,0}=1$ and where the $F_l$ functions are defined as
\begin{align}
&F_0(n,m) = \frac{1}{4}(m+\frac{1}{2})^3 q^3-\frac{3}{4} (n+\frac{1}{2})(m+\frac{1}{2})^2 q^2 b+\frac{3}{4} (n+\frac{1}{2})^2 (m+\frac{1}{2})qb^2 -\frac{1}{4} (n+\frac{1}{2})^3 b^3 \nonumber \\
 & - P^2 \Bigl [(m+\frac{1}{2}) q  - (n+\frac{1}{2}) b \Bigr]\\
&F_1(n,m) =-(m+\frac{1}{2})(\frac{3}{4}m^2 +\frac{3}{2}m+ \frac{13}{16} )q^3 +  \frac{3}{2} (n+\frac{1}{2}) (m+\frac{1}{2})(m+1)  q^2 b  -\frac{3}{4}(m+\frac{1}{2}) (n+\frac{1}{2})^2 qb^2\nonumber \\
&+  P^2(m+\frac{1}{2})  q \\
&F_2(n,m)= \frac{3}{4} (m+\frac{1}{2})(m+\frac{3}{2})^2 q^3 -\frac{3}{4}(n+\frac{1}{2})(m+\frac{1}{2})(m+\frac{3}{2})bq^2 \\
&F_3(n,m)= -\frac{1}{4} (m+\frac{1}{2})(m+\frac{3}{2})(m+\frac{5}{2})q^3\, .
\end{align}
Since the recursion for the Gelfand-Dikii coefficients is \emph{one-step}, using formula~\eqref{I2n-1} and~\eqref{GDrecFinalLiouville} is a very efficient way of computing the $I_{2n-1}$, which have also been checked numerically by exploiting TBA equation (\ref{LiouvilleTBA}). Besides, we have repeated the calculations in the case of the minimal models and have found the same formul\ae \, in terms of $c$ and $\Delta$ (as expected).

For $b=1$ the recursion formula \eqref{GDrecFinalLiouville} simplifies hugely, so that we can write closed formul\ae \, for the self-dual Liouville LIM 
\be
I_{2n-1} (b = 1, P^2)= \sum_{k=0}^n  \Upsilon_{n,k} P^{2k}
\label{LIM-b=1}
\ee
at any $n$, as polynomials in $P^2$ with the first leading terms: $\Upsilon_{n,n} = (-1)^n$, $\Upsilon_{n,n-1}=  (-1)^n \frac{1}{12}n(n-\frac{1}{2})$, $\Upsilon_{n,n-2} = (-1)^n \frac{1}{1440} (n-1)n (n-\frac{1}{2})(7n - \frac{3}{2})$\footnote{Hasmik Poghosyan has solved \eqref{GDrecFinalLiouville} for general $b$.}.

\section{Baxter's $T$ and $Q$ functions at self-dual point as Seiberg-Witten cycles} \label{TandQ}
 

This section is devoted at the $b=1$ case, where we first analyse an important connexion between $T$ and the Floquet exponent, as anticipated numerically by \cite{ZamolodchikovMemorial}. Then, we give both $T$ and $Q$ two peculiar SW theory interpretations. As anticipated, in the self-dual GME \eqref{GMF1}, we shall rotate the real into the imaginary axis, $z = - i y-\pi$, and obtain the Mathieu equation
\begin{equation}
-\frac{d^2}{dz^2} \psi(z,\theta) + \Bigl [ 2 e^{2\theta} \cos{z}  - P^2 \Bigr] \psi(z, \theta) = 0 \label{MathieuODE/IM} \, . 
\end{equation}
According to Floquet theorem, there exist two linearly independent (quasi-periodic) solutions of the Mathieu equation~\eqref{MathieuODE/IM} of the form $\psi_+(z)=e^{ \nu z} p(z)$ and $\psi_-(z)=e^{-  \nu z} p(-z)$, with periodic $p(z) = p(z + 2 \pi)$ and monodromy exponent $\nu= \nu(\theta, P)$, \emph{the Floquet index}. Al. B. Zamolodchikov conjectured that the cosine of the Floquet index is equal to the Baxter's $T$ function for the self dual Liouville model $b=1$
\begin{equation}
T(\theta, P^2) = 2 \cosh \bigl \{2 \pi \nu(\theta, P^2) \bigr\}   \label{ZamConj} \,\, .
\end{equation}
We can give some hints on the reason of this relation upon looking at the $TQ$ relation~\eqref{TQ} at $b=1$
\begin{equation}  \label{TQ1}
T(\theta) = \frac{Q(\theta + i \pi/2)}{Q(\theta)} +\frac{Q(\theta - i \pi/2)}{Q(\theta)} 
\end{equation}
where in the r.h.s. there are these wronskians ({\it cf.} \eqref{QW} {\it et seq.}) $Q(\theta) = W[U_0,V_0](\theta)$, $Q(\theta \pm i \pi /2) = W[U_{\pm 1},V_0](\theta) $, all expressible in the Floquet basis. Nevertheless, we will leave the proof to another occasion since this identity has a very relevant interpretation in gauge theory once we add the other important ingredient, namely the coincidence of the quantum SW cycle (\ref{NScycles}) $a= - i \nu$ with the Floquet exponent. More precisely, the Mathieu ODE/IM equation~\eqref{MathieuODE/IM} coincides with the Seiberg-Witten one~\eqref{SWeq}, provided we set the change of variables
\begin{align}
\frac{\hslash }{\Lambda}&= e^{-\theta }\,\, ,\qquad \frac{u}{\Lambda^2}  = \frac{P^2}{2 e^{2 \theta}}.      \label{GaugeIntID}
\end{align}
Thus, the above~\eqref{ZamConj} can be interpreted as a direct connexion between the Baxter's $T$ function and the quantum SW cycle (\ref{NScycles}):
\begin{align}
T(\hslash,u,\Lambda) \equiv T(\theta, P^2 ) &=2 \cos \left \{  2 \pi a( \hslash,  u,\Lambda) \right \} \label{idTa} \,\, .
\end{align}

Now we find an analogous link for the $Q$-function, $Q(\theta,P^2)$, upon writing \eqref{GMF1} in the gauge variables \eqref{GaugeIntID}
\begin{align} \label{mMathieuSU2}
-\frac{\hslash^2 }{2} \frac{d^2}{dy^2} \psi(y) +[ \Lambda^2 \cosh{y} + u] \psi(y)&=0 \,\, ,
\end{align}
which is the same as equation \eqref{SWeq} upon substitution $\psi(y)=\psi(z)$ with $y=i z+i \pi$. Equation \eqref{mMathieuSU2} gives rise for $\CP(y)=-i \frac{d}{dy}\ln \psi(y)$ to the Riccati equation
\be
\mathcal{P}^2(y, \hslash, u ) -i \frac{d\mathcal{P}(y, \hslash, u)}{dy} =-( \frac{2u}{\hslash^2} + \frac{2\Lambda^2}{\hslash^2} \cosh{y}) \label{RicPy}, 
\ee
while $\mathcal{P}(z)=- i \frac{d}{dz}\ln \psi(z)$ (so that $\mathcal{P}(y) dy = \mathcal{P}(z) dz$) verifies 
\be
\mathcal{P}^2(z, \hslash, u) - i \frac{d\mathcal{P}(z ,\hslash, u)}{dz} =  \frac{2u}{\hslash^2}- \frac{2\Lambda^2}{\hslash^2} \cos{z} \label{RicPz}.
\ee
We fix the solution of \eqref{mMathieuSU2} to be $V_0(y,\hslash,u)$ \eqref{V0S} (which the Baxter $Q$ \eqref{Xexp}) by choosing the sign $\CP_{-1}(y)=-i \sqrt{\frac{2u}{\hslash^2} + \frac{2\Lambda^2}{\hslash^2} \cosh{y}}$ so that we reproduce the SW cycle \eqref{a0D} by integrating $\CP_{-1}(z)=\sqrt{\frac{2u}{\hslash^2}- \frac{2\Lambda^2}{\hslash^2} \cos{z}}$, namely the $z$-quasi-periodic solution $\psi_+(z, \hslash, u)$. In detail, the solution $\CP(y)$ is meromorphic with simple poles (repeating periodically: $\CP(y+ 2 \pi i) = \CP(y)$), corresponding to the zeroes of $V_0(y)$. As $\hslash \to 0$, the poles become denser and denser and form the branch cut characteristic of the semi-classical expansion. Anyway, the poles are on the line $\Im y = \pi$ and on the segment $\Re y = 0$, $ \arccos (- u/\Lambda^2)= \pi - \arccos (u/\Lambda^2) < \Im y < \pi +\arccos (u/\Lambda^2)$ and repeat $2\pi i$-periodically. Indeed, these are also the branch cuts of the semi-classical expansion. Since $\ln Q$ \eqref{Xexp} is $i$ times the integral over $(-\infty,+\infty)$ of the regularised NS momentum \eqref{Preg} (as $b=1$) 
\be
\CP_{reg}(y)= \CP(y) +2 i e^{\theta}\cosh \frac{y}{2} - \frac{i}{4} \tanh y \,\, ,
\label{Preg'}
\ee 
let us consider the integral of $i \CP_{reg}(y)$ on the (oriented) closed curve which runs along the real axis, slightly below the cut and closes laterally. Mathematically, it is $\gamma = \gamma_1 \cup \gamma_{lat,R} \cup \gamma_2 \cup  \gamma_3 \cup \gamma_4 \cup \gamma_5 \cup \gamma_{lat,L}$, with $\gamma_1 =(-\infty , +\infty)$, $\gamma_2 = (+\infty + i \pi - i 0  , 0^+ + i \pi - i 0)$ , $\gamma_3 = (0^+ +  i \pi  - i 0, 0^+ +i \pi - i\arccos (u/\Lambda^2)$, $\gamma_4 = (0^- + i \pi - i \arccos (u/\Lambda^2), 0^- + i \pi - i 0 )$, $\gamma_5 = (0^- + i \pi - i 0  , -\infty + i \pi -i 0 )$, and $\gamma_{lat,L}$ $\gamma_{lat,R}$ are the lateral contours which close the curve. As all the poles are outside the contour, this integral is zero. Let us now consider the different pieces. As recalled, the integral on $\gamma_1$ equals $\ln Q$. The integrals on the lateral contours $\gamma_{lat,L/R}$ are (exponentially) suppressed (since $\mathcal{P}_{reg}(y) = O(e^{\mp y/2})$ for $ \Re{y} \to \pm \infty$). For $t\in \mathbb{R}$, $\CP(t + i \pi -i0) =-\CP(-t + i \pi -i0)$ as they verify the same Riccati equation \eqref{RicPy} and share the same 
$\CP_{-1,I}(t + i \pi -i0) =-\CP_{-1,I}(-t + i \pi -i0)$, where the subscript $I$ means that they are both evaluated on the first sheet of the Riemann surface. From this we easily deduce (as the hyperbolic functions in \eqref{Preg'} have no cut and behave in odd manner under $t\to -t$)
\be
\CP_{reg}(t -i0 + i \pi) =-\CP_{reg}(-t -i0 + i \pi) \, ,\,\,\, t\in \mathbb{R} \,\,\, ,\label{Pregodd} 
\ee
and, as a consequence, that the integrals on $\gamma_2$ and $\gamma_5$ cancel each other. In conclusion, the only remaining integrals are those on $-\gamma_3$ and $-\gamma_4$, which do not have contribution from the hyperbolic functions (without cuts). They can be better taken into account in the variable $z$ with the property $\CP(-z+i0) =-\CP(z-i0)$ for $z$ real so that they are respectively the two following terms 
\be
\int_{-\arccos (u /\Lambda^2)} ^{0}  \CP(z-i0) \, dz +  \int_{0}^{-\arccos (u /\Lambda^2)}  \CP(z+i0) \, dz  =\int_{-\arccos (u /\Lambda^2) -i0}^{ +\arccos (u /\Lambda^2) -i0}  \CP(z ) \, dz = \int_{-\infty }^{+\infty } \mathcal{P}_{reg}(y) \, dy \,.  
\ee
The last equality entails the connexion of the self-dual Liouville Baxter's $Q$ function with the dual cycle~\eqref{NScycles}:
\be
Q(\theta , P^2 )\equiv Q(\hslash,u,\Lambda) = \exp \Bigl \{ 2 \pi \i a_D( \hslash,u,\Lambda) \Bigr \} \label{idQaD} 
 \ee
Strictly speaking, we have proven this for the range $0<u<\Lambda^2$ ($\hslash$ real), but we can immagine to extend the relation by analytic continuation.

In consideration of the one to one relation between $\theta$ and $\hslash$ \eqref{GaugeIntID} we can use the first in place of the latter. Thus, the  $\Re{\theta} \rightarrow +\infty$ (small $\hslash$) asymptotic expansions of $T(\theta)$ and $Q(\theta)$ in the strip $| \Im{\theta} | < \frac{\pi}{2} + \epsilon$, $\epsilon>0$, are 
\begin{align}
T(\theta, P^2 )= T( \theta ,u) 
&\doteq 2 \cos \biggl \{  2\pi  \sum_{n=0}^\infty e^{\theta(1-2n)}\Lambda^{2n-1}     a^{(n)}(u,\Lambda) \biggr \} \qquad  \label{NSTexp} \\
Q(\theta ,P^2)= Q(\theta , u) 
&\doteq  \exp \biggl \{ 2\pi i \sum_{n=0}^\infty  e^{\theta(1-2n)}  \Lambda^{2n-1} a^{(n)}_D(u,\Lambda) \biggr \} \qquad   \label{NSQexp} \, .
\end{align}

We now find a new way to compute the NS-deformed Seiberg Witten cycles modes, which will also reveal itself to be an asymptotic check of the identification~\eqref{NSQexp}. Considering the large energy asymptotic expansion~\eqref{QexpI2n-1} of $Q$ in terms of the LIM, we observe that, since in Seiberg Witten theory $u$ is finite as $\theta \to +\infty$, it is necessary that also $P^2(\theta) =  2\frac{u}{\Lambda^2} e^{2 \theta} \to +\infty$. In this double limit, an infinite number of LIMs $I_{2n-1}(b=1)$ are re-summed into an NS-deformed dual cycle mode (a sort of charge in its turn). Then the $n$-th mode of the $Q$ function in the small $\hslash$ expansion~\eqref{NSQexp} is a series which gives the $n$-th dual cycle 
\begin{align}
 2 \pi i a^{(n)}_D(u,\Lambda) &=  - \Lambda^{1-2n}\sum_{k=0}^\infty  2^k C_{n+k}   \Upsilon_{n+k,k} \Bigl( \frac{u}{\Lambda^2}\Bigr)^k  \label{aDnLIM}\,.
\end{align}
From here, closed formul\ae \, can be obtained through the previous powerful method for determining the LIM ~\eqref{LIM-b=1}; they are very simple series (cf.~\eqref{operatorB_nm}) convergent in the circle $|u| < \Lambda^2$:
\begin{align}
2 \pi i a^{(0)}_D(u,\Lambda) &= -\Lambda \sum_{n=0}^\infty \biggl [ (-1)^n 2^n \frac{\Gamma^2(\frac{n}{2} -\frac{1}{4})}{4 \sqrt{\pi} n!} \biggr ]  \Bigl( \frac{u}{\Lambda^2}\Bigr)^n\label{aD0usmall}\\
2 \pi i a^{(1)}_D(u,\Lambda) &= \Lambda^{-1}\sum_{n=0}^\infty  \biggr [(-1)^n 2^n \frac{(n+\frac{1}{2})\Gamma^2(\frac{n}{2} +\frac{1}{4})}{48 \sqrt{\pi} n!}  \biggr ] \Bigl( \frac{u}{\Lambda^2}\Bigr)^n\\
2 \pi i a^{(2)}_D(u,\Lambda) &= -\Lambda^{-3} \sum_{n=0}^\infty \biggl [ (-1)^n 2^n \frac{(n+\frac{3}{2})(7n + \frac{25}{2})  \Gamma^2(\frac{n}{2} +\frac{3}{4})}{5760 \sqrt{\pi} n!} \biggr ]  \Bigl( \frac{u}{\Lambda^2}\Bigr)^n \, .
\end{align}
Conversely, we can invert \eqref{aDnLIM} and expresses the LIMs in terms of the the deformed cycles. Therefore, thanks to the quantum Picard-Fuchs equations (\ref{PF0}-\ref{PF2}), we can express explicitly the LIM  themselves \eqref{LIM-b=1} at all orders.

\section{Functional relations, gauge TBA and $\mathbb{Z}_2$ symmetry} \label{functional}

As we have a gauge interpretation~\eqref{idTa} and \eqref{idQaD} of the self-dual Liouville integrability Baxter's $T$ and $Q$ functions, respectively, we can search for a gauge interpretation of the integrability functional relations (the $QQ$ system, the $TQ$ relation, the periodicity relation, \emph{cf.} Section~\ref{Zam} with $b=1$). First, we write the $QQ$ relation~\eqref{QQZ} at $b=1$, and then the same in the gauge variables \eqref{GaugeIntID}
\begin{equation}
1 + Q^2(\theta,P^2) = Q(\theta - i \pi /2,P^2) Q(\theta + i \pi /2,P^2)\, , \qquad 1 + Q^2(\theta,u) = Q( \theta - i \pi /2, -u) Q( \theta + i \pi /2, -u) \, , \label{gaugeQQ}
\end{equation}
where we have considered that $\theta \to \theta \mp \i \pi /2$ means $u\to -u$ (as $P^2$ is fixed). The latter equation, the gauge $QQ$ system, has been verified by using the expansion \eqref{NSQexp} in several complex regions of $u$, in particular in the circle $|u|< \Lambda^2$. In the present case it is a 'square root' of the $Y$ system and then gives us the gauge TBA equations. In fact, we can take the logarithm of both members and invert to obtain an explicit expression for $\ln Q(\theta,u)$. As usual, this inversion possesses zero-modes and so does not fix completely the forcing term. For it we need to consider the asymptotic expansion \eqref{NSQexp} as $\Re \theta \to + \infty$, $\ln Q(\theta,u)  \simeq  2 \pi i a^{(0)}_D(u,\Lambda) e^{\theta}/\Lambda$. In this way we find a TBA integral equation for the deformed dual cycle $-2 \ln Q(\theta,u) =  \varepsilon (\theta,u) = -4 \pi i  a_D( \hslash(\theta) ,u)$ and then write the same for $u\rightarrow -u$
\begin{align} 
\begin{split}
   \varepsilon(\theta, u,\Lambda)   &=  -4 \pi i  a^{(0)}_D(u,\Lambda)  \frac{e^{\theta}}{\Lambda}  - 2 \int_{-\infty}^\infty \frac{ \ln \left [ 1 +  \exp \{ - \varepsilon(\theta', -u,\Lambda) \}  \right] }{\cosh{(\theta- \theta')}} \frac{d \theta'}{ 2 \pi}    \\
    \varepsilon(\theta, -u,\Lambda)   &= - 4 \pi i  a^{(0)}_D(-u,\Lambda)  \frac{e^{\theta}}{\Lambda}  - 2 \int_{-\infty}^\infty \frac{ \ln \left [ 1 +  \exp \{ - \varepsilon(\theta', u,\Lambda) \}  \right] }{\cosh{(\theta- \theta')}} \frac{d \theta'}{ 2 \pi} \label{gaugeTBA} \, .
    \end{split}
\end{align}
In contrast with Liouville TBA (where was no $P$), the forcing terms have non-trivial $u$-dependences, the SW cycles indeed, which can be interpreted (as in \cite{GaiottoMooreNeitzke:2008}) as the mass of a BPS state of a monopole and dyon (via Bilal-Ferrari formul\ae, {\it i.e.} \eqref{anDananD} for $n=0$), respectively. Actually, the quantum cycle 
\begin{align}
2 \pi i a_D( \hslash(\theta) , - u,\Lambda)  &=  2 \pi i a^{(0)}_D(- u,\Lambda)  \frac{e^{\theta}}{\Lambda}  + \int_{-\infty}^{\infty} \frac{ \ln \left [ 1 +  \exp \left \{ 4 \pi i a_D(\hslash(\theta') ,u,\Lambda) \right \} \right ] }{\cosh{(\theta- \theta')}} \frac{d \theta'}{ 2 \pi}  \,\, . \label{cycleTBA}
\end{align}
can take the place of the first cycle $a(\hslash, u)$ (linked to $T$ in any case) as the latter can be expressed in terms of the former two via \eqref{TQ}. From the large $\theta$ asymptotic expansion of the integral part, we find a also all the NS-deformed dual cycles modes ($m \geq 1$)
\begin{align}
2 \pi  i\, a^{(m)}_D(u, \Lambda) &= -\Lambda^{1-2m}(-1)^m \int_{-\infty}^\infty e^{\theta' (2m-1)}\ln \Bigl [  1 + \exp \{ - \varepsilon(\theta', -u,\Lambda)) \}  \Bigr ]  \frac{d \theta'}{  \pi} \label{aDnTBA} \, .
\end{align}
By solving with numerical iterations the two coupled equations of gauge TBA \eqref{gaugeTBA}, we tested these expressions with the analytic WKB recursive cycles (\ref{operatorB_nm},\,\ref{recB_nm}) for a region of the complex plane slightly larger than $|u|< \Lambda^2$. The $u=0$ unique equation from \eqref{gaugeTBA} was conjectured numerically in ~\cite{GaiottoOpers}.

Consider now the $TQ$ relation~\eqref{TQZ} at $b=1$, which we also write in the gauge variables \eqref{GaugeIntID}
\begin{equation}
T(\theta, P^2 ) = \frac{Q(\theta - i \pi /2, P^2)+Q(\theta + i \pi /2, P^2)}{Q(\theta, P^2)} \, ,\qquad  T( \theta,u)  = \frac{Q(\theta - i \pi/2,  -u ) + Q(\theta + i \pi/2,  -u )}{Q( \theta , u) }
 \label{TQ}
\end{equation}
For the asymptotic $\hslash \to 0$ analysis of the latter relation, we keep only the dominant exponents (fixed by SW order \eqref{aD0usmall})  
\begin{equation}
\exp \Bigl \{ -\sgn{(\Im{u})}  2 \pi i \sum_{n=0}^\infty e^{\theta(1-2n)}  a^{(n)}(+u) \Bigr \} \doteq 
\exp \Bigl \{  - 2 \pi \sum_{n=0}^\infty e^{\theta(1-2n)} \Bigl[ \sgn{(\Im{u})} (-1)^n a^{(n)}_D(-u) +i  a^{(n)}_D(u) \Bigr] \Bigr \}  . 
\end{equation}
Thus, the $TQ$ relation entails 
\begin{equation}
a_D^{(n)}(-u) =  i (-1)^n\left[ \sgn{(\Im{u})} \,a_D^{(n)}(u) - a^{(n)}(u)\right ] .\label{anDananD}
\end{equation}
These relations are, in fact, the extension of the $\mathbb{Z}_2$ symmetry relation in SW ($n=0$) \cite{BilalFerrari:1996} to the NS-deformed theory \cite{BasarDunne:2015}. In a nutshell, the $TQ$ relation encodes such $\mathbb{Z}_2$ relations among the asymptotic modes as a unique exact equation. Relation~\eqref{anDananD} allows one to express the NS-cycles completely in terms of the NS-dual cycles. Thus, the new formulas~\eqref{aDnLIM}~\eqref{aDnTBA} we have found for $a^{(n)}_D(u)$, actually determine also $a^{(n)}(u) = -\sgn{(\Im{u})} \,a_D^{(n)}(u)+ i (-1)^n a_D^{(n)}(-u)$. 

We finally consider the integrability~\eqref{TperZ} $T$ periodicity relation at $b=1$: 
\begin{equation}
T(\theta,P^2 ) = T(\theta- i \pi /2,P^2)  \qquad T( \theta , u)  = T(\theta - i \pi/2, -u ) \label{Tper}
\end{equation}
To interpret this relation through the asymptotic identification~\eqref{NSTexp}. Thus, the~\eqref{Tper} relation truncates to
\begin{align}
 \exp \Bigl \{ -\sgn{(\Im{u})} 2 \pi  i \sum_{n=0}^\infty e^{\theta(1-2n)} a^{(n)}(u) \Bigr \} \doteq  \exp \Bigl \{ + 2 \pi  \sum_{n=0}^\infty e^{\theta(1-2n)} (-1)^n a^{(n)}(-u) \Bigr \}
\end{align}
from which, we deduce the $\mathbb{Z}_2$ symmetry relation for the other cycle~\cite{BilalFerrari:1996} extended to the NS-deformed theory~\cite{BasarDunne:2015}
\begin{equation}
a^{(n)}(-u) = - i (-1)^n \sgn{(\Im{u})}    \, a^{(n)}(u) \, . \label{anan}
\end{equation}
We conclude that, thanks to the identifications~\eqref{idTa}~\eqref{idQaD} between the integrability and gauge quantities, we can interpret the Baxter's $TQ$ relation~\eqref{TQ} and $T$ periodicity relation~\eqref{Tper} as \emph{non-perturbative} $\mathbb{Z}_2$ symmetry relations.

\section{Theories with matter and perspectives}
\label{conclusions-matter}
As we will motivate in future, if the gauge theory possesses massless $N_f=1$ flavours, the null-vector equation is that of a Liouville theory with $b=1\sqrt{2}$, while $N_f=2$ (still chiral) brings again $b=1$. Chiral $N_f=3$ should involve again $b=1$. Notice that the relation with integrability involves always $T$ and $Y$ (not $Q$) as, for instance, the three node TBA of $N_f=1$ proves. But still the possible r\^ole of $b$ needs to be better investigated and understood. 

Higher rank gauge groups keep the parallel (as one can see from the $SU(3)$ case).

In conclusion, the powerful ODE/IM correspondence has been revealing a very suggestive connexion between the quantum integrable models and $\epsilon_1$-deformed Seiberg-Witten $\mathcal{N}=2$ supersymmetric gauge theories. And in this sense the correspondence yields a natural quantisation scheme for SW theory. A latere, an efficient asymptotic expansion technique is presented: a one-step recursion for the computation of the WKB asymptotic expansions of the wave function both for large energy and small Planck constant, or, in other words, the local integrals of motion of Liouville CFT and the expansion modes of the deformed Seiberg-Witten cycles. Eventually, they are related each other by our gauge/integrability link. And must satisfy all orders Pichard-Fuchs equations. It would be also interesting to explore the implications for the cycles as described in~\cite{BourgineFioravanti:2017A}~\cite{BourgineFioravanti:2017B}.

The r\^ole of the R-symmetry and its breaking in connexion with integrability deserve more investigations, but a simple parallel we can put forward is the similarity with planar $\mathcal{N}=4$ SYM, where the residual symmetry and integrability are used to construct the spectrum \cite{FioravantiPiscagliaRossi:2015}.

Besides the long time thinking about these topics (at least since the collaboration \cite{BourgineFioravanti:2017A}, see also \cite{dan-thesis}), the sending out of the present work has been today prompted by the appearance of the paper \cite{GrassiGuMarino} which may overlap with ours.

\textbf{Acknowledgments.}
We are particularly indebted to R. Poghossian and H. Poghosyan for many important suggestions, numerical integrations and checks (especially by mean of instanton calculus). Moreover we would like to thank P. Dorey, M.L. Frau, F. Fucito, S. Lukyanov, F. Morales, M. Rossi, A. Lerda R. Tateo, D. Masoero for stimulating discussions. This project was partially supported by the grants: GAST (INFN), the MPNS--COST Action MP1210, the EC Network Gatis and the 2019 PRIN. D.F. thanks the Galileo Galilei Institute for Theoretical Physics (GGI) for invitation to the workshop 'Supersymmetric Quantum Field Theories in the Non-perturbative Regime'.

\bibliographystyle{utphys}

\providecommand{\href}[2]{#2}\begingroup\raggedright\endgroup

\end{document}